\documentclass[letter]{aa}

\usepackage{graphicx}
\usepackage{txfonts}
\usepackage{booktabs}
\usepackage[normalem]{ulem}
\usepackage{multirow}
\usepackage{hyperref}
\usepackage{orcidlink}
\begin{document} 

   \title{A Triple AGN in the NGC\,7733--7734 Merging Group}


   \author{Jyoti Yadav \orcidlink{0000-0002-5641-8102}
          \inst{1,}\inst{2}\thanks{jyoti [at] iiap.res.in}
          \and
          Mousumi Das \orcidlink{0000-0001-8996-6474} \inst{1}
          \and
          Sudhanshu Barway \orcidlink{0000-0002-3927-5402} \inst{1}
          \and
          Francoise Combes \orcidlink{0000-0003-2658-7893} \inst{3}}

   \institute{Indian Institute of Astrophysics, Koramangala II Block, Bangalore 560034, India\\
  \and
Pondicherry University, R.V. Nagar, Kalapet, 605014, Puducherry, India\\
\and
Observatoire de Paris, LERMA, College de France, CNRS, PSL University, Sorbonne University, F-75014 Paris, France}


 
  \abstract
   {Galaxy interactions and mergers can lead to supermassive black hole (SMBH) binaries which become active galactic nuclei (AGN) pairs when the SMBHs start accreting mass. If there is a third galaxy involved in the interaction, then a triple AGN system can form.}
   {Our goal is to investigate the nature of the nuclear emission from the galaxies in the interacting pair NGC\,7733--NGC\,7734 using archival VLT/MUSE Integral field spectrograph data and study its relation to the stellar mass distribution traced by near-infrared (NIR) observations from the South African Astronomical Observatory (SAAO).
   }
   {We conducted near-infrared observations using the SAAO and identified the morphological properties of bulges in each galaxy. We used MUSE data to obtain a set of ionized emission lines from each galaxy and studied the ionization mechanism. We also examined the relation of the galaxy pair with any nearby companions with Far-UV observations using the UVIT.}
   { The emission line analysis from the central regions of NGC\,7733 and NGC\,7734 show Seyfert and LINER type AGN activity. The galaxy pair NGC\,7733--34 also shows evidence of a third component, which has Seyfert-like emission. Hence, the galaxy pair NGC\,7733--34 forms a triple AGN system. We also detected an Extended Narrow-line region (ENLR) associated with the nucleus of NGC\,7733.}
  {}

   \maketitle
%
\section{Introduction}
Galaxy interactions and mergers are the major drivers of galaxy evolution in our low redshift universe, leading to the growth of supermassive black holes (SMBHs), bulges, and massive galaxies \citep{DiMatteo2005Natur.433..604D}. One of the most favorable environments for such activity are galaxy groups where galaxies are closely interacting, especially those that have significant reservoirs of cold gas that can be used to fuel star formation and active galactic nuclear (AGN) activity \citep{Eastman2007ApJ...664L...9E, Georgakakis2008MNRAS.391..183G, Arnold2009ApJ...707.1691A, Martini2009ApJ...701...66M, Tzanavaris2014ApJS..212....9T}. 

The tidal forces during galaxy interactions may trigger the formation of bars and spiral arms that produce gravitational torques on the stars and gas in the galaxy disks. These, in turn, drive gas into the nuclear regions, triggering central star formation \citep{Springel2005ApJ...620L..79S, Hopkins2011MNRAS.415.1027H}. The gas accretion onto the SMBHs can trigger AGN activity, but the details of how the gas reaches the inner kpc region are still not clear. However, processes such as nuclear spirals and kpc-scale bars likely play an important role \citep{combes.2001}. Observations suggest that most galaxies are associated with groups or have some companions \citep{McGee2009MNRAS.400..937M}. The galaxy interactions within the group can strongly affect the galaxy morphologies, and trigger star formation and AGN activity if the galaxies are gas-rich \citep{duc.renaud.2013}. Gas can also be pulled out due to tidal interaction and ram pressure stripping, as seen in the tails of HI gas or star-forming knots in H$\alpha$ and ultraviolet (UV) \citep{Kenney2004AJ....127.3361K,  Linden2010MNRAS.404.1231V, Yadav2021arXiv210316819Y}.

\begin{figure*}
    \centering
    \includegraphics[width=0.98\textwidth]{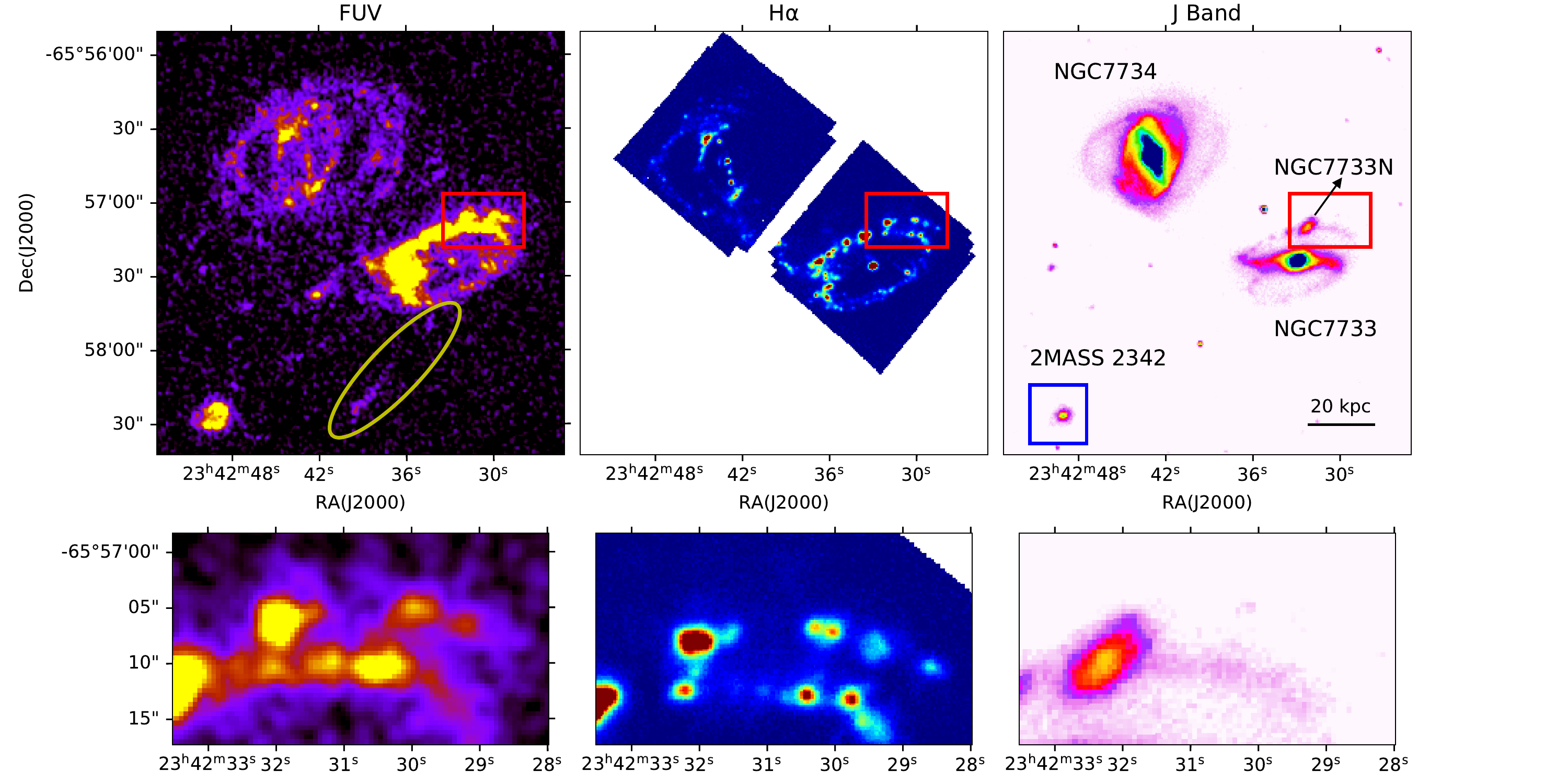}
        \caption{Upper panel shows the FUV, H$\alpha$ and J band images of galaxy group NGC\,7733--34. Bottom panel shows the zoomed in version of the red rectangles. Blue box represents the possible merging galaxy 2MASS\,2342. Yellow ellipse shows the bridge between NGC\,7733 and merger candidate 2MASS\,2342.}
    \label{fig:images}
\end{figure*}

As galaxies merge, the dynamical friction on the SMBHs in the individual galaxies causes them to move closer together. If the SMBHs are accreting, they can form AGN pairs \citep{rubinur.etal.2019}.
Studies show that there is a higher fraction of dual AGN in interacting galaxies, which indicates that galaxy interactions can trigger AGN activity \citep{2014Satyapal,2015Kocevski,2018Koss}. Several dual AGN have been detected locally \citep{2012Koss, rubinur.etal.2018, Imanishi2020ApJ...891..140I}, and at higher redshifts \citep{2008Myers, 2010Hennawi, Silverman2020ApJ...899..154S, Silva2021ApJ...909..124S}. However, only a few binary AGN have been detected \citep{kharb.etal.2017}.

Galaxy interactions can also lead to triple merger systems, and if the SMBHs of the individual galaxies are accreting, it will form a triple AGN system. Although such systems are rare, a few have been discovered \citep{2012Koss, Pfeifle2019ApJ...883..167P}. It is important to detect more multiple AGN, in order to understand how AGN activity can affect the merging processes on kpc scales. In this letter, we present the detection of a kpc-scale triple AGN system in a nearby galaxy pair NGC\,7733--NGC\,7734 (hereafter NGC\,7733--34). The instruments, methods and results are described below. The value of Hubble constant (H$_{0}$) used in this paper is 67.8 km s$^{-1}$ Mpc$^{-1}$.

\begin{table}
    \centering
    \small 
    \begin{tabular}{lcccr}
    \toprule
    Source & R.A & Dec. & z  \\
    & (hh mm ss)& (dd mm ss) &    \\
    \hline
    NGC\,7733 & 23:42:32.79 & -65:57:27.42 & 0.03392$\pm$0.00003 \\ 
    NGC\,7734 & 23:42:43.20 & -65:56:44.19 & 0.03527$\pm$0.00006\\
    NGC\,7733N   & 23:42:32.25 & -65:57:09.43 & 0.03619$\pm$ 0.00012 \\
    2MASS\,2342 & 23:42:49 & -65:58:26 &  \\
    \toprule
    \end{tabular}
    \caption{Details of the galaxies. Column 2 and 3 are the coordinates of galaxies in J2000.0. Column 4: redshift of NGC\,7733 and NGC\,7734 taken from \citet{1996Mathewson} and \citet{1991deVaucouleurs} respectively.}
    \label{tab:details}
\end{table} 

\section{NGC\,7733--34 Group}
NGC\,7733--34 is a group of interacting galaxies at a distance of 154 Mpc \citep{1987Arp}. NGC7734 is
at a distance of 48.3 kpc and has a velocity of 406 km s$^{-1}$ with respect to NGC7733 \citep{1991deVaucouleurs}.
NGC\,7733 is a barred spiral that shows knots in the arms and hosts a Seyfert 2 nucleus \citep{2016Tempel}. NGC\,7734 is a barred spiral with peculiar arms \citep{1991deVaucouleurs, 1989Paturel}).
The atomic gas content of the galaxy pair is $\leqslant 4.2 \times 10^{9}$ M\textsubscript{\(\odot\)} and the molecular content of NGC\,7733 and NGC\,7734 is $\leqslant 15.4 \times 10^{8}$ M\textsubscript{\(\odot\)} and $\simeq 1.073 \times 10^{10}$ M\textsubscript{\(\odot\)} respectively \citep{1997Horellou}. \citet{2001Jahan-Miri} identified the star-forming knots and their ages in the individual galaxies and found that NGC\,7733 has a much younger stellar population than NGC\,7734. The properties of the galaxies are given in Table~\ref{tab:details}.

\section{OBSERVATIONS AND DATA ANALYSIS}
\subsection{UV DATA}
We observed NGC\,7733--34 in the far-ultraviolet (FUV) band using the Ultraviolet Imaging Telescope (UVIT) on board ASTROSAT \citep{2012Kumar}. The UVIT has two telescopes that can simultaneously observe in three bands FUV (1300--1800 \AA), NUV (2000--3000 \AA), and the visible band. Drift correction were done with the visible channel. The UVIT has a spatial resolution of around 1\arcsec\ and a field of view of 28$\arcmin$. We reduced the UVIT level 1 data downloaded from the Indian Space Science Data Centre (ISSDC) using CCDLAB \citep{2017PASP..129k5002P,2020PASP..132e4503P}.

\begin{figure*}
   \centering
   \includegraphics[width=0.98\textwidth]{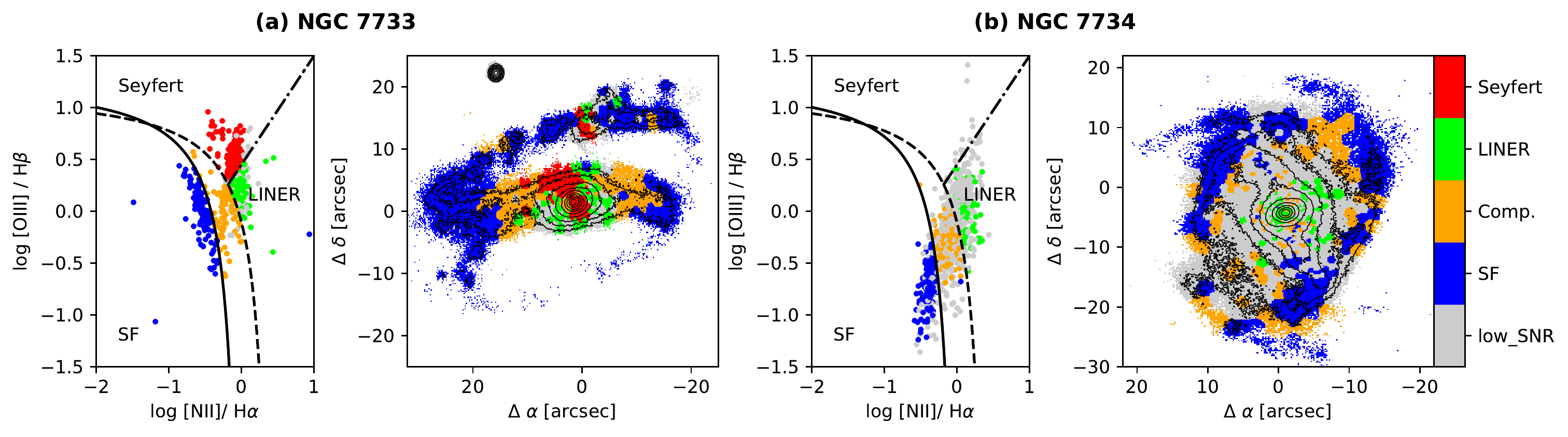}
        \caption{ [\ion{N}{ii}] BPT Diagrams of NGC\,7733 and NGC\,7734.
        $\Delta \alpha$  and $\Delta \delta$ are R.A. and declination with respect to the central position of the source, which is indicated in Table~\ref{tab:details}. The solid curve in the BPT is taken from \citet{2003Kauffmann}, and the region below this line shows the ionization due to star-forming regions.
        The dashed curve is taken from \citet{2001Kewley}, and the region above this line shows the ionization due to AGN. The dashed-dotted line is taken from \citet{2007Schawinski}, which divides the LINER and Seyfert AGN.
        The bins with an amplitude-to-noise (A/N) ratio of less than 4 in any of the lines are shown in gray.}
        \label{fig:bpt_whan_ngc7733_34}%
    \end{figure*}

\subsection{Near Infrared DATA}
We performed deep Near Infrared (NIR) imaging of the galaxy group NGC\,7733--34 with the SIRIUS camera \citep{2003SPIE.4841..459N} on the Infrared Survey Facility (IRSF) 1.4 m telescope in Sutherland, South Africa. The SIRIUS camera has three 1024$\times$1024 HgCdTe detectors, which give simultaneous imaging in the J, H, and K bands, with a field of view 7.7\arcmin\ $\times$ 7.7\arcmin, and the pixel scale is 0.45\arcsec\ \citep{2012Nagayama}. The images were taken in automatic dithering mode with $\sim$20\arcsec\ steps with individual frame exposure times of 30 seconds each, giving a total exposure time of typically $\sim$120 minutes for the combined set of exposures. Data reduction was carried out with a pipeline for the SIRIUS observations, including corrections for non-linearity, dark subtraction, and flat fielding. 

\subsection{MUSE DATA}
We have used Multi Unit Spectroscopic Explorer (MUSE) archival data to study the optical emission lines from NGC\,7733--34. MUSE is an integral field spectrograph (IFU) on the Very Large Telescope (VLT) and gives 3D spectroscopic data cubes with very high resolution. We used the data from the wide-field mode, which has a field of view of 1\arcmin\ $\times$ 1\arcmin. It has a spatial sampling of 0.2\arcsec\ $\times$ 0.2\arcsec\ and a spectral resolution of 1750 at 4650{\AA} to 3750 at 9300\AA.

We ran the Galaxy IFU Spectroscopy Tool (GIST\footnote{\url{https://abittner.gitlab.io/thegistpipeline/}} version 3.0.3; \citealt{2019Bittner}) pipeline to analyze the MUSE data. GIST uses a python implemented version of GANDALF \citep{2006Sarzi, 2006Falcon, 2019Bittner} for fitting the gas emission lines and Penalized Pixel-Fitting (pPXF) \citep{2004Cappellari,2017Cappellari} method for stellar continuum fitting. GIST gives stellar kinematics and gas emission line properties based on binned data over a given wavelength range and signal-to-noise (S/N). 
For stellar velocity calculation, the code convolves the linear combination of spectral energy templates with line of sight velocity distribution and fits that with the spectra from each bin.

We have Voronoi binned the data based on H$\alpha$ (6555--6575 \AA) emission. For binning, we used an S/N of 30. We have also corrected the spectra for the Milky Way extinction. We adopted the method used in \citet{2021comeron} and fitted the continuum using a multiplicative eighth-order Legendre polynomial. We used a minimum S/N of 3 to remove noisy data before binning the data.
We have used GIST emission line fluxes for [\ion{O}{iii}], [\ion{N}{ii}], H$\alpha$ and H$\beta$ for further analysis.

Fig.~\ref{fig:images} shows the FUV, H$\alpha$ and NIR J band images of the galaxy group. The FUV and J band images reveal an extended source (2MASS\,23424898-6558257 referred to as 2MASS\,2342) southeast of the galaxy NGC\,7733. The FUV image reveals a bright bridge in the southeastern region connecting NGC\,7733 to 2MASS\,2342. The extended object in the southeastern region of NGC\,7733 could be interacting with NGC\,7733 and forms part of the NGC\,7733--34 group.
The projected spatial separation of 2MASS\,2342 from NGC\,7733 is $\sim$ 87 kpc assuming that both galaxies are at the same redshift.

\section{Nature of the emission from the nuclei in NGC\,7733--34}
To study the mechanism producing the ionized gas, we derived the BPT plots of the two galaxies \citep{1981Baldwin}. The BPT diagram uses the
[\ion{O}{iii}]$\lambda 5007/$H$\beta$ and the [\ion{N}{ii}]$\lambda 6563/$H$\alpha$ line ratios to classify the origin of the emission. 
Fig.~\ref{fig:bpt_whan_ngc7733_34} (a) and (b) show the BPT plots of NGC\,7733 and NGC\,7734 respectively. 
The nuclear emission from NGC\,7733 lies in the Seyfert part of the BPT plot, while NGC\,7734 shows LINER emission from the central region. 
The BPT diagram of NGC\,7733 also shows the ongoing star formation in the arms and Seyfert-like emission along the minor axis of the bar.

\begin{figure}
    \centering
    \includegraphics[width=0.45\textwidth]{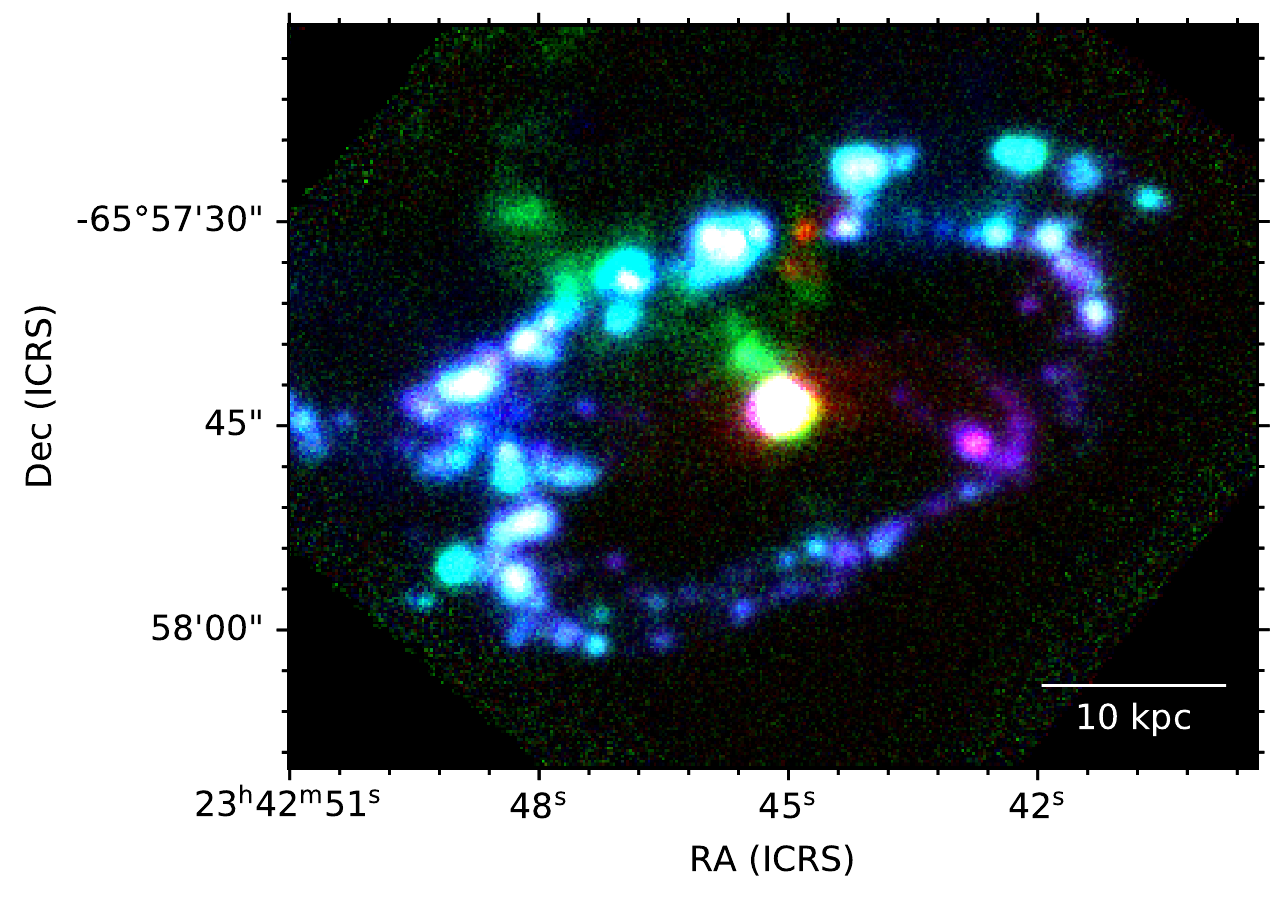}
    \caption{Color composite Image Of NGC\,7733. Blue, green, and red represent the H$\alpha$, [\ion{O}{iii}] and [\ion{N}{ii}] emissions respectively. The [\ion{O}{iii}] emission towards the northeastern side shows the ENLR.}
    \label{fig:Ha_OIII_NII}
\end{figure}

Fig.~\ref{fig:Ha_OIII_NII} shows the color composite image of H$\alpha$ (blue), [\ion{O}{iii}] (green) and [\ion{N}{ii}] (red) emissions, respectively. The bright H$\alpha$ emission, in blue along the arms of NGC\,7733, confirms the ongoing recent star formation in the arms of the galaxy. The [\ion{O}{iii}] emission is extended towards the northeast, which confirms that it is due to emission from the extended narrow-line region (ENLR) along the minor axis of the galaxy. The orientation is due to projection effects. The ENLR is associated with photoionized gas around the AGN and can be traced out to $\sim$18 kpc in the northeastern region of NGC\,7733. The velocity field and the trailing spirals suggest that the NE side is the near side of the galaxy, so the [\ion{O}{iii}] emission is extended towards us. The southeastern ENLR is extended away from our line of sight and is not visible due to obscuration by the dust in the galaxy. Ongoing recent star formation activity in NGC\,7733 \citep{2001Jahan-Miri} suggests the presence of gas and dust in this galaxy.

\section{Is there a third galaxy?}
\begin{figure}
    \centering
    \includegraphics[width=0.4\textwidth]{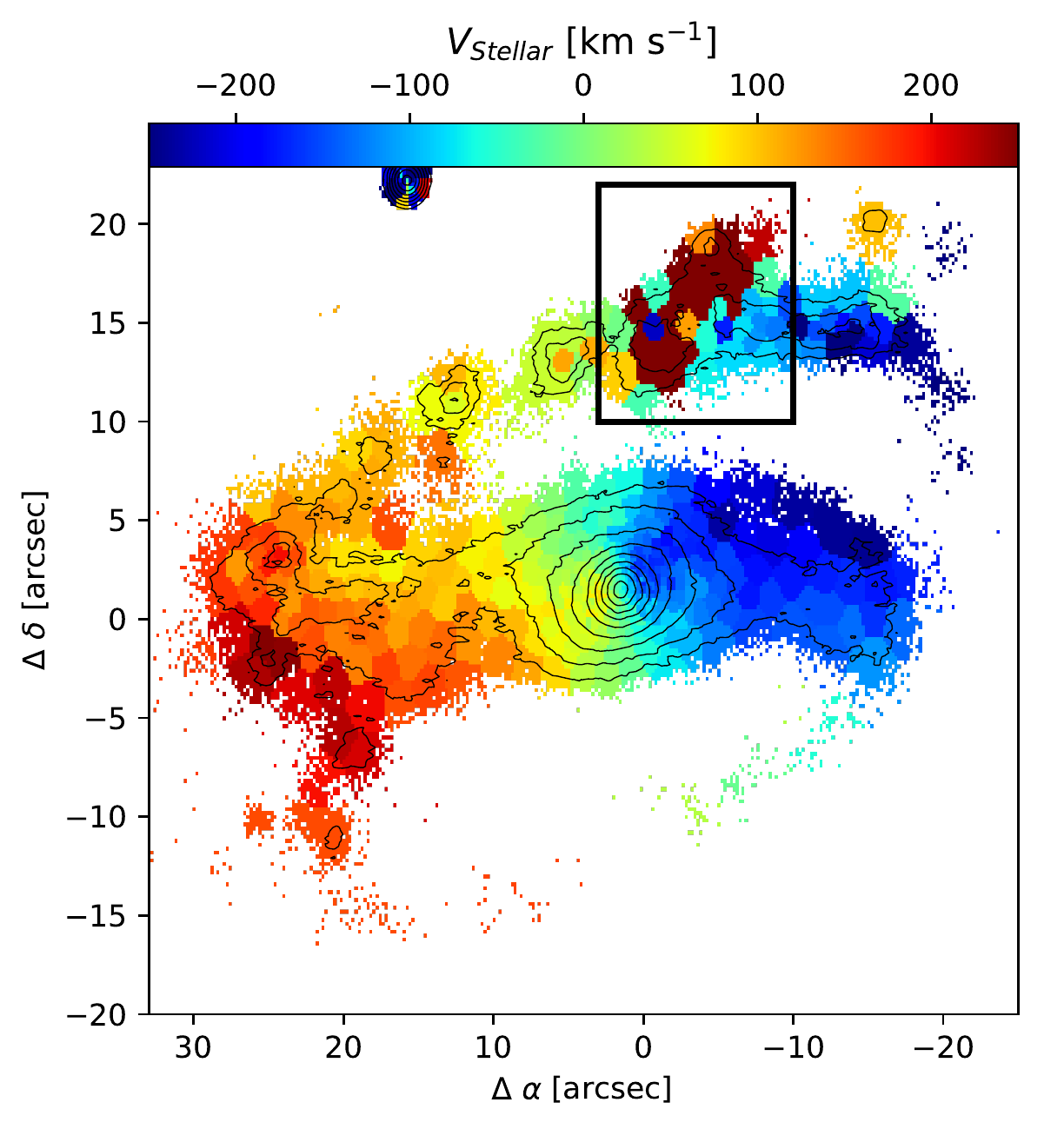}
    \caption{Stellar velocity map of NGC\,7733 obtained from GIST. $\Delta \alpha$  and $\Delta \delta$ are R.A. and Declination with respect to the central position of the source, which is indicated in Table~\ref{tab:details}. The black rectangle in the northwest shows that the velocity of NGC\,7733N is different from galactic rotation. }
    \label{fig:velocity_33}
\end{figure}

The FUV, J band, and MUSE H$\alpha$ images show a large bright knot in the northern arm of NGC\,7733. This region was identified as a star-forming knot in \citet{2001Jahan-Miri}. The knot appears to be extended in the J band as shown in the red box in Fig.~\ref{fig:images}.
Fig.~\ref{fig:velocity_33} shows the velocity field of the galaxy NGC\,7733. Most of the velocity field shows features typical of a rotating disk, with the Doppler shifted velocities of the approaching (blue) and receding sides (yellow and red). However, the velocity of the knot (shown in the rectangle) is clearly different from the velocity field of the disc.

\subsection{Decomposition using GALFIT}
We decomposed the NIR images of NGC\,7733, and NGC\,7734 into  bulge, disc, and bar components using GALFIT \citep{2002AJ....124..266P} 2-D decomposition code. We modeled both the bulges and the bars using Sersic profiles, and the discs with exponential profiles. A $\chi^2$ minimization technique was used to carry out this fit. We also used the Sersic profile to fit the knot on the northwest arm of NGC\,7733 using GALFIT. We found that the effective radius r$_e$ of the Sersic profile in the $J$ band is 3.204\arcsec\ which corresponds to $\sim$2.4 kpc. Such a large value for the radius of the knot suggests that it is likely to be the bulge of a galaxy, instead of being a star-forming region within NGC\,7733. 

\subsection{Emission line Analysis of the Knot in NGC\,7733}
\begin{figure}
    \centering
    \includegraphics[width=0.49\textwidth]{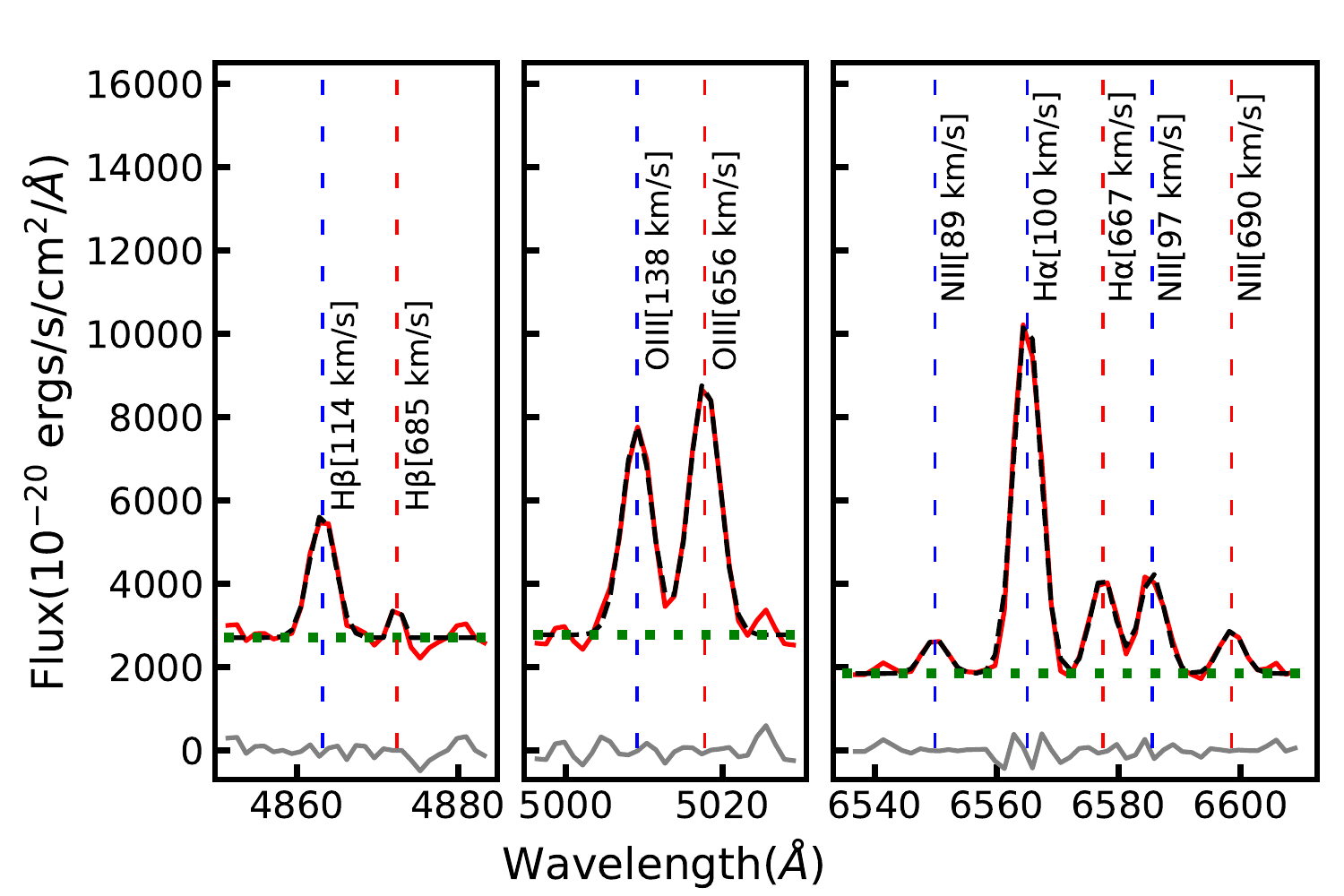}
    \caption{ The blue and red vertical dashed lines correspond to the spectra of NGC\,7733 and NGC\,7733N, respectively. Observed and fitted spectra are shown in red and black. Continuum is shown by dotted green horizontal lines. The vertical dashed lines show the peaks of emission lines along with their respective velocities.}
    \label{fig:emission_line_fitting}
\end{figure}

The knot shows a different velocity compared to the northern arm of NGC\,7733 as indicated by the rectangle in Fig.~\ref{fig:velocity_33}. The MUSE spectra from this region show that there are two sets of emission lines, as shown in Fig.~\ref{fig:emission_line_fitting}. The knot appears to be redshifted by $\sim$650 km s$^{-1}$ with respect to the galaxy. This is a strong indication that the knot is not a part of the galaxy and is instead a separate system that is visually overlapping with the arm of the galaxy. We will use `NGC\,7733N' to refer to the knot in the following text.
To check the nature of emission from NGC\,7733N, we used the fluxes of H$\beta$, [\ion{O}{iii}], H$\alpha$ and [\ion{N}{ii}] emission lines for all bins in the rectangle shown in Fig.~\ref{fig:velocity_33}. We divided the spectrum into three wavelength ranges, 4851--4883, 4996--5029, and 6532--6608 \AA. We fitted Gaussian profiles to H$\beta$, [\ion{O}{iii}], H$\alpha$ and [\ion{N}{ii}] lines and a constant continuum as shown in Fig.~\ref{fig:emission_line_fitting}. The emission lines from some of the bins could not be fitted due to poor S/N. The resultant fits confirmed that NGC\,7733N has its own set of emission lines.
We calculated the mean velocity of H$\alpha$ in NGC\,7733 and NGC\,7733N after fitting the emission lines. We found that NGC\,7733N is moving with a mean velocity of +655$\pm$46 km s$^{-1}$ with respect to NGC\,7733 and is at a redshift of 0.03619$\pm$ 0.00012.

\begin{figure}
   \centering
     \includegraphics[width=0.35\textwidth]{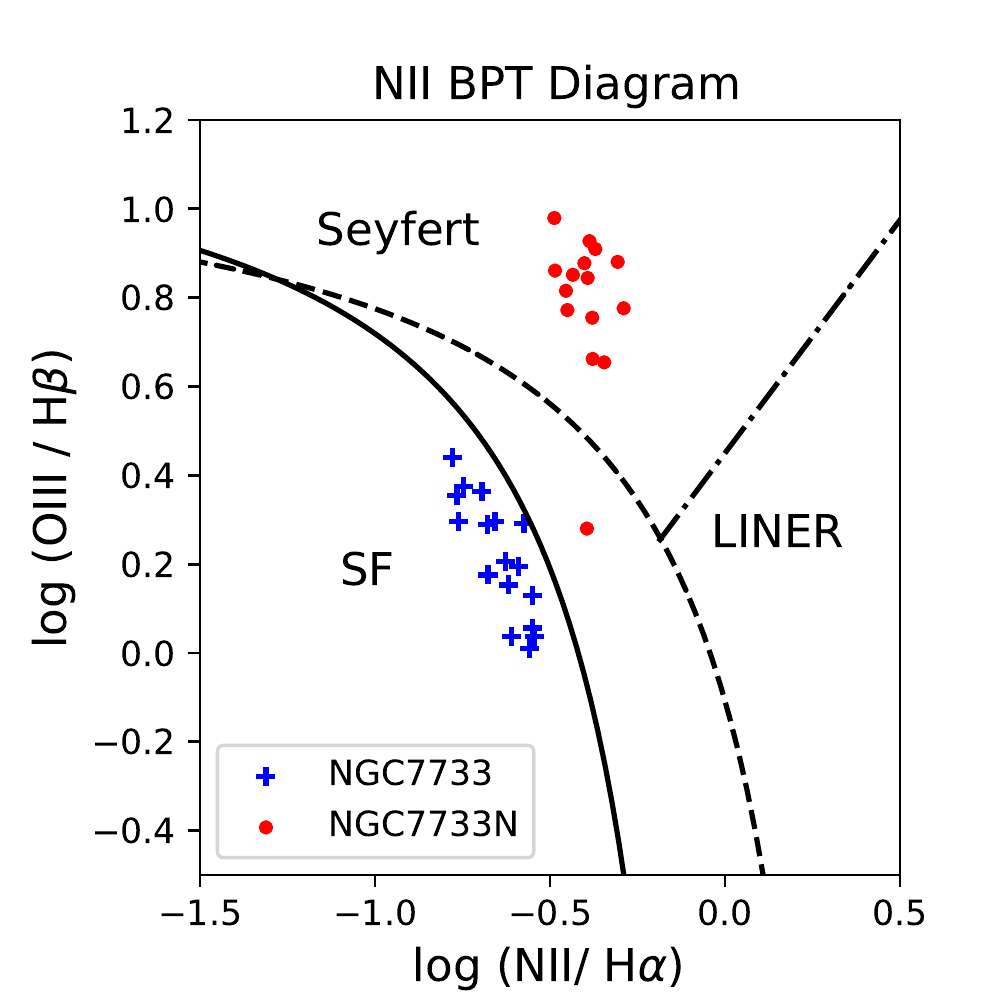}
     \caption{BPT diagram for NGC\,7733 and its companion NGC\,7733N. Blue crosses and red dots color represents the emission from NGC\,7733 and NGC\,7733N respectively.}
     \label{fig:BPT_whan_33_33A}%
    \end{figure}

The BPT diagram of two sets of emission lines is shown in Fig.~\ref{fig:BPT_whan_33_33A}. The red markers and blue crosses show the emission from NGC\,7733N, and NGC\,7733, respectively. The emission from the NGC\,7733 component of this region is due to the star formation, which is expected since it originates from the arm. The emission from NGC\,7733N lies in the Seyfert region, indicating an AGN in this galaxy.

\section{Is this a triple system?}
\subsection{Tidal features in UV and H$\alpha$}

Fig.~\ref{fig:images} shows an arm-like structure above the main arm in the northern part of NGC\,7733 in the UVIT FUV and MUSE H$\alpha$ images. The arm or tidal tail starts from the western side and merges with the location of NGC\,7733N. It is likely to have formed via the tidal interaction between NGC\,7733 and NGC\,7733N. There are UV and H$\alpha$ bright knots in the tidal arm, which suggests that there is ongoing massive star formation triggered by the interaction. NGC\,7733N then represents the bulge of the merging companion galaxy (Fig.~\ref{fig:images}), which may have already lost most of its disk mass in the tidal interaction.

\begin{table*}
    \centering
    \footnotesize
    \begin{tabular}{lccccccccccr}
    \toprule
    Galaxy	& 	 M$_K$	&	Re	& log(M$_{\star}$) &	\multicolumn{8}{c}{Black Hole Mass (10$^{8}$ M$_{\odot}$)} \\
    \cmidrule(lr){5-12}
	&	(mag)	&	 (arcsec)	&   &	\multicolumn{2}{c}{M$_{\bullet}$--L$_K$}			&	\multicolumn{3}{c}{M$_{\bullet}$--$\sigma$}					&	\multicolumn{3}{c}{M$_{\bullet}$-M$_{bulge}$}					\\
	\cmidrule(lr){5-6} \cmidrule(lr){7-9} \cmidrule(lr){10-12}
	&		&	 &	&	MH2003	&	KH2013	&	Gu2009	&	MM2013	&	KH2013	&	MH2003	&	MM2013	&	KH2013	\\
	\hline \\
    NGC\,7733	&	-24.09	&	1.73	& 11.39 &	1.83$^{+0.62}_{-0.43}$	&	4.68$^{+1.14}_{-1.03}$	&	0.54$^{+0.05}_{-0.05}$	&	0.64$^{+0.03}_{-0.03}$	&	1.24$^{+0.06}_{-0.07}$	&	1.31$^{+0.89}_{-0.31}$	&	1.86$^{+0.73}_{-0.56}$	&	3.04$^{+0.91}_{-0.73}$	\\ \\
    NGC\,7734	&	-23.74	&	1.58	& 11.81 &	1.28$^{+0.36}_{-0.26}$ &	
    3.18$^{+0.90}_{-0.77}$	&
    1.13$^{+0.20}_{-0.18}$	&
    1.70$^{+0.18}_{-0.17}$	&
    2.63$^{+0.28}_{-0.26}$	&	
    0.90$^{+0.60}_{-0.21}$	&
    1.25$^{+0.57}_{-0.41}$	&
    1.98$^{+0.67}_{-0.52} $	\\ \\
    NGC\,7733N	&	-22.11	&	3.1	& 9.96 &	0.23$^{+0.01}_{-0.01}$	&
    0.51$^{+0.24}_{-0.17}$	& 
    &		&  	&	
    0.15$^{+0.10}_{-0.04}$	&
    0.19$^{+0.14}_{-0.09}$	&
    0.26$^{+0.14}_{-0.09}$	\\ \\
    \toprule
    \end{tabular}
    \caption{Properties of Galactic Bulges and central black holes. Column 2: K band Absolute magnitude of bulges. Column 3: Effective radius of bulges. Column 4: Stellar mass estimated in K band using $M/L = 0.6$ \citep{2014McGaughAJ....148...77M}. Column 5 and 6: Mass of central black holes using M$_{\bullet}$ and K band luminosity.
    Column 7, 8 and 9: Mass of central black holes using M$_{\bullet}$--$\sigma$ relation.
    Column 10, 11 and 12: Mass of central black holes using M$_{\bullet}$-- Mass of bulge relation.
    MH2003: \citet{Marconi2003ApJ...589L..21M}, KH2013: \citet{Kormendy2013ARA&A..51..511K}, {Gu2009: \citet{Gultekin2009ApJ...698..198G}}, {MM2013: \citet{McConnell2013ApJ...764..184M}}.}
    \label{tab:masses}
\end{table*}

\subsection{Masses of the central Black holes in NGC\,7733, NGC\,7733N and NGC\,7734}
We estimated the K band magnitude and effective radius of the bulges in NGC\,7733, NGC\,7733N, and NGC\,7734 using GALFIT and corrected the magnitude for Galactic extinction using \citet{Schlafly2011ApJ...737..103S}.
We estimated the stellar mass (M$_{\star}$) of each galaxy using Mass--Luminosity relation \citep{2014McGaughAJ....148...77M}. Based on the stellar mass ratios, NGC\,7733-NGC\,7734 is a major while NGC\,7733N-NGC\,7733 is a minor merging system, respectively.
The nuclear black hole masses (M$_{\bullet}$) were estimated using the following relations; (i)~M$_{\bullet}$--luminosity in the K band \citep{Marconi2003ApJ...589L..21M, Kormendy2013ARA&A..51..511K}, (ii)~the M--$\sigma$ relation \citep{Gultekin2009ApJ...698..198G, McConnell2013ApJ...764..184M, Kormendy2013ARA&A..51..511K} and the (iii)~M$_{\bullet}$--bulge mass relation \citep{Marconi2003ApJ...589L..21M, McConnell2013ApJ...764..184M, Kormendy2013ARA&A..51..511K}. The SMBH masses for the three AGN and the stellar masses of the galaxies are listed in Table~\ref{tab:masses}. The bulge velocity dispersion of NGC\,7733N is not reliable, so we could not calculate the black hole mass in NGC\,7733N using the M--$\sigma$ relation.
The nuclear black hole masses of NGC7733 (0.54--4.68 $\times10^8$ M$_{\odot}$) and NGC7734 (0.90--3.18 $\times10^8$ M$_{\odot}$) have a similar range of values, and their stellar masses are also comparable. NGC\,7733N hosts the least massive stellar disk and the least massive nuclear black hole (1.53--5.15 $\times10^7$ M$_{\odot}$) in the triple AGN system.

\section{Implications for our triple AGN detection}
Dual AGN have been detected in many galaxies, but triple AGN are rare. Our study shows that multiple AGN systems can be present in galaxy groups that show ongoing mergers. Since galaxy groups are fairly common \citep{tempel.etal.2017}, they can be good targets for detecting multiple AGN. Another important implication of our study is the enhanced AGN feedback in such systems,  which can heat the intra-cluster medium in clusters and enrich the CGM around galaxies \citep{Fabian2012ARA&A..50..455F,tumlinson.etal.2017}. Although  multiple AGN feedback is relatively unexplored, it must play an important role in building hot gas reservoirs around galaxies, especially at early epochs when galaxy mergers were more frequent \citep{Lackner2014AJ....148..137L, Man2016ApJ...830...89M, Mundy2017MNRAS.470.3507M}. Feedback processes such as jets or intense quasar outflows from AGN can affect the evolution of the host galaxy \citep{Salom2016A&A...595A..65S, Harrison2018NatAs...2..198H}. Multiple AGN enhance feedback effects may lead to faster SMBH growth \citep{volonteri.etal.2020}. There may be additional star formation if the AGN outflows interact with each other. Triple AGN systems such as the one we have detected in NGC\,7733--34 are thus ideal laboratories to study the growth of hot gas, AGN feedback, and SMBH growth in mergers and should be relatively common in gas-rich galaxy groups.

\section{Conclusions}
\begin{enumerate}
    \item We find that the paired galaxies NGC\,7733 and NGC\,7734 host Liner and Seyfert type nuclei, respectively.
    \item We also detected an ENLR from the AGN in NGC\,7733 towards the northeastern region.
    \item The multiwavelength data shows a tidal arm above the northern arm of NGC\,7733, and the tidal arm connects with the location of a third galaxy which we call NGC\,7733N. 
    \item We have confirmed the existence of the third galaxy NGC\,7733N in the NGC\,7733--34 group. It appears to overlap with the northern arm of NGC\,7733. The emission line analysis of NGC\,7733N shows that it hosts a Seyfert nucleus. 
    \item The UV and NIR images show a fourth galaxy (2MASS\,2342) lying southeast of the galaxy pair, which could be part of the merging system. We see a potential bridge between NGC\,7733 and 2MASS\,2342 in the UV map. This bridge is composed of hot gas that has been pulled out during the interaction phase. However, redshift estimation is necessary to confirm this.
\end{enumerate}
Triple-AGN are rare, however, detailed studies could enhance their observed frequency in groups, and especially at high redshift. Although this study focuses only on one system, the results suggest that small merging groups are ideal laboratories to study multiple AGN systems. 
    
\begin{acknowledgements}
We thank the anonymous referee for the thoughtful
review, which improved the impact and clarity of this work. This publication uses the data from the UVIT, which is part of the AstroSat mission of the Indian Space Research Organisation (ISRO), archived at the Indian Space Science Data Centre (ISSDC). This research has also used NIR data from IRSF at the South African Astronomical Observatory (SAAO). Part of the results are based on observations collected at the European Southern Observatory under ESO programme 0103.A-0637, run B. This research has made use of the NASA/IPAC Extragalactic Database (NED) which is operated by the Jet Propulsion Laboratory, California Institute of Technology, under contract with the National Aeronautics and Space Administration. M.D. acknowledges the support of Science and Engineering Research Board (SERB) MATRICS grant MTR/2020/000266 for this research.
\end{acknowledgements}

%
%
\bibliographystyle{aa}
\bibliography{sample}
\onecolumn

\end{document}